\documentclass[aps,prl,twocolumn, superscriptaddress,groupedaddress]{revtex4}  
\usepackage{graphicx}  
\usepackage{dcolumn}   
\usepackage{bm}        
\usepackage{amssymb}   

\def\be{\begin{equation}}
\def\ee{\end{equation}}
\newcommand{\bea}{\begin{eqnarray}}
\newcommand{\eea}{\end{eqnarray}}

\newcommand{\hv}{\langle H_u^0\rangle}
\newcommand{\dn}{\Delta N_{\rm eff}}
\newcommand{\ntilde}{\tilde{N}_1}
\newcommand{\mn}{m_{N_1}}
\newcommand{\mntilde}{m_{\tilde{N}_1}}
\newcommand*\diff{\mathop{}\!\mathrm{d}}

\begin{document}

\widetext
\leftline{MCTP-16-26}

\title{Sterile Neutrino Dark Matter with Supersymmetry}

\author{Bibhushan~Shakya}
\affiliation{Department of Physics, University of Cincinnati, Cincinnati, OH 45221, USA}
\affiliation{Michigan Center for Theoretical Physics, University of Michigan, Ann Arbor, MI 48109, USA}
\author{James D. Wells} 
\affiliation{Michigan Center for Theoretical Physics, University of Michigan, Ann Arbor, MI 48109, USA}
\affiliation{Deutsches-Elektronen Synchrotron (DESY), Notkestra\ss e 85, Hamburg, D-22607, Germany}

\vskip 0.25cm

\begin{abstract}
Sterile neutrino dark matter, a popular alternative to the WIMP paradigm, has generally been studied in non-supersymmetric setups. If the underlying theory is supersymmetric, we find that several interesting and novel dark matter features can arise. In particular, in scenarios of freeze-in production of sterile neutrino dark matter, its superpartner, the sterile sneutrino, can play a crucial role in early Universe cosmology as the dominant source of cold, warm, or hot dark matter, or of a subdominant relativistic population of sterile neutrinos that can contribute  to the effective number of relativistic degrees of freedom $N_{\text{eff}}$ during Big Bang nucleosynthesis. 
\end{abstract}

\maketitle

\section{Motivation}
\label{sec:motivation}

A sterile neutrino is a well motivated and widely studied dark matter (DM) candidate. 
In the Neutrino Minimal Standard Model ($\nu$MSM) \cite{Asaka:2005an,Asaka:2005pn,Asaka:2006nq}, its relic abundance is produced through its mixing with the active neutrinos via the Dodelson-Widrow mechanism \cite{Dodelson:1993je} for keV scale masses; however, this possibility has now been ruled out by a combination of X-ray and Lyman-alpha measurements \cite{Boyarsky:2006fg,Boyarsky:2006ag, Boyarsky:2005us,Boyarsky:2007ay,Boyarsky:2007ge,Seljak:2006qw, Asaka:2006nq, Boyarsky:2008xj,Horiuchi:2013noa}. The Shi-Fuller mechanism \cite{Shi:1998km} employs resonant production, but requires fine-tuned parameters and faces constraints from structure formation \cite{Horiuchi:2015qri,Schneider:2016uqi}. Thermal freeze-out with additional interactions followed by appropriate entropy dilution can also result in the correct relic abundance \cite{Bezrukov:2009th,Nemevsek:2012cd,Shuve:2014doa,Patwardhan:2015kga}, but is strongly constrained by Big Bang nucleosynthesis \cite{King:2012wg}. 

An alternate production mechanism that is compatible with all constraints is the freeze-in mechanism \cite{Chung:1998rq,Hall:2009bx}, where the relic abundance is built up through a feeble coupling to some particle beyond the Standard Model (BSM) present in the early Universe. This possibility has been studied by several groups in several motivated frameworks \cite{Shaposhnikov:2006xi,Petraki:2007gq,Kusenko:2006rh,McDonald:1993ex,Yaguna:2011qn,Merle:2013wta,Adulpravitchai:2014xna,Kang:2014cia,Roland:2014vba,Roland:2015yoa,Frigerio:2014ifa,Kadota:2007mv,Abada:2014zra} (see \cite{Shakya:2015xnx} for a recent summary). While the details differ, all of these frameworks share two common salient features: 

(1) a vanishing mixing between the sterile neutrino DM candidate $N_1$ and the active neutrinos, necessary to make $N_1$ stable or very long-lived and  to alleviate tension with observations, and 

(2) a feeble coupling between $N_1$ and a BSM particle present in the early Universe, which facilitates DM production. 

It should be noted that (1) may appear unnatural at face value, but can be rendered technically natural in the limit of a $\mathcal{Z}_2$ symmetry that $N_1$ is charged under, which could be built into the details of the underlying model. 

Studies of sterile neutrino DM in the literature are generally performed in non-supersymmetric setups. However, independent of dark matter considerations, there are several compelling reasons to expect the underlying theory of nature to be supersymmetric. The purpose of this paper is to study a supersymmetric extension of the sterile neutrino dark matter framework with properties (1) and (2) above, which are generic, model-independent features of the freeze-in mechanism. In this framework, $N_1$ is part of a supermultiplet that also contains a scalar, the sterile sneutrino $\tilde{N}_1$. The aforementioned $\mathcal{Z}_2$ symmetry necessarily requires $\ntilde$ to decay into $N_1$; furthermore, as we will see, this decay involves the ``feeble" coupling from (2) above, hence $\ntilde$ can potentially be long-lived. These features allow for interesting modifications of early Universe cosmology and dark matter properties. 

\section{Framework}

The sterile neutrino DM freeze-in framework requires the following Lagrangian terms \cite{Shaposhnikov:2006xi,Petraki:2007gq,Kusenko:2006rh,McDonald:1993ex,Yaguna:2011qn,Merle:2013wta,Adulpravitchai:2014xna,Kang:2014cia,Roland:2014vba,Roland:2015yoa,Frigerio:2014ifa,Kadota:2007mv,Abada:2014zra,Shakya:2015xnx,Merle:2015oja} (we only list terms that will be relevant for our study):
\be
\mathcal{L}\supset y_{ij} L_i h N_j+x_i \phi \bar{N}^c_i N_i + \lambda (H^\dagger H) \phi^2.
\label{eq:lagrangian}
\ee
In addition to the three Standard Model (SM)-singlet, sterile neutrinos $N_i$, this setup also features a neutral scalar $\phi$. $x, y$ are dimensionless numbers. The aforementioned requirement of vanishing mixing for $N_1$ translates to $y_{i1}\rightarrow 0$, corresponding to a $\mathcal{Z}_2$ symmetry for $N_1$. The second term leads to freeze-in production of $N_1$ via $\phi\rightarrow N_1N_1$ decays if the coupling $x_1$ is ``feeble", $x_1^2\,\textless\,m_\phi/M_{Pl}$ \cite{Hall:2009bx}, where $M_{Pl}$ is the Planck mass. If $\phi$ obtains a vacuum expectation value, this term also gives rise to Majorana masses for the sterile neutrinos; we do not consider this possibility here. Finally, the third term accounts for the SM interactions of $\phi$ necessary for its presence in the early Universe. 

In a supersymmetric theory, each of the above fields is part of a supermultiplet; we denote the supermultiplets as $\Phi$ and ${\mathcal{N}}_i$, with their spin $(0,1/2)$ components being ($\phi,\psi$) and ($\tilde{N}_i,N_i$) respectively. The Lagrangian terms in Eq.\,\ref{eq:lagrangian} can then be generated from the following superpotential:
\be
\label{eq:newterms}
W\supset y_{ij} \mathcal{L}_i H_u \mathcal{N}_j+x_i \Phi \mathcal{N}_i\mathcal{N}_i + \sqrt{\lambda} \Phi H_u H_d\,. 
\label{eq:superpotential}
\ee
This superpotential further generates the following additional terms (we only list the ones that will be relevant for our study):
\be
\mathcal{L}\supset x_i \psi N_i \tilde{N}_i + \sqrt{\lambda}\phi\tilde{H}_u\tilde{H}_d 
+\sqrt{\lambda}(\psi h_d\tilde{H_u}+\psi h_u\tilde{H_d})
\label{eq:lagrangian2}
\ee
In addition, the following soft terms are also generated after SUSY breaking:
\be
\mathcal{L}_{soft}\supset y_{ij}{A_y}_{ij}\tilde{L}_i h_u \tilde{N}_j+ x_i{A_x}_i \phi\ntilde\ntilde+\sqrt{\lambda}A_\lambda \phi h_u h_d.
\label{eq:soft}
\ee
Note, in particular, that the second term can give rise to the decay $\phi\rightarrow\ntilde\ntilde$.

In keeping with previous work on freeze-in of sterile neutrino dark matter \cite{Shaposhnikov:2006xi,Petraki:2007gq,Kusenko:2006rh,McDonald:1993ex,Yaguna:2011qn,Merle:2013wta,Adulpravitchai:2014xna,Kang:2014cia,Roland:2014vba,Roland:2015yoa,Frigerio:2014ifa,Kadota:2007mv,Abada:2014zra, Shakya:2015xnx}, we take $N_1$ to be light (sub-GeV scale). $N_2, N_3$ are taken to be above the GeV scale to ensure they decay before BBN and remain compatible with cosmological constraints. The heavier particles $\phi,\,\psi,$ and $ \tilde{N}_i$ are at some heavy scale $m_{\text{SUSY}}$, the scale of supersymmetry. For concreteness, we also assume R-parity and take the lightest supersymmetric particle (LSP) to be a sub-TeV Higgsino, which therefore makes up a small fraction of dark matter.  

\begin{figure}[t!]
\includegraphics[width=2.7in]{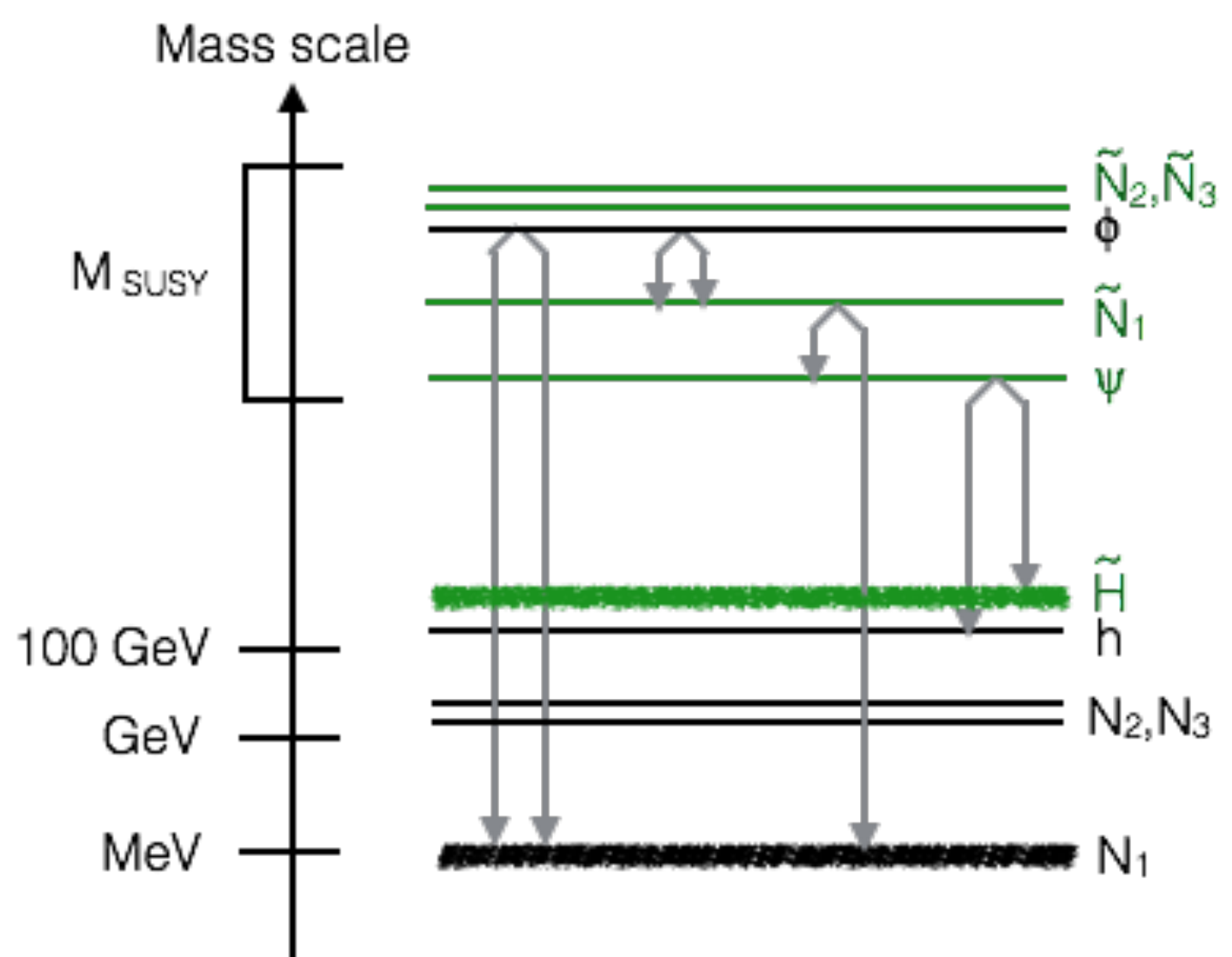}
\caption{\label{fig:spectrum} Particle masses and relevant decays. Supersymmetric particles are shown in green to highlight how the non-supersymmetric sterile neutrino freeze-in framework gets extended. Particles that make up dark matter are denoted by think lines.}
\end{figure}

In general, several permutations of particle masses and couplings are possible. In this paper, we take $m_\phi\,\textgreater\,m_{\ntilde}$, so that $\phi$ decays into both the dark matter candidate $N_1$ and its superpartner $\ntilde$ (this is not strictly necessary for $\ntilde$ production, as $\ntilde$ also gets produced through annihilation processes in the early Universe). The $\mathcal{Z}_2$ symmetry forces $\ntilde$ to necessarily decay into $N_1$, and therefore through the $x_1\psi N_1 \ntilde$ operator (see Eq.\,\ref{eq:lagrangian2}). If $m_{\ntilde}\,\textgreater\,m_\psi$, $\ntilde$ decays as $\ntilde\rightarrow \psi N_1$. Otherwise, if $m_{\ntilde}\,\textless\,m_\psi$, the decay occurs either through an off-shell $\psi$ as $\ntilde\rightarrow N_1\tilde{H}h$, or as $\ntilde\rightarrow \tilde{H} N_1$ through $\psi-\tilde{H}$ mixing, induced by the final term in Eq.\,\ref{eq:lagrangian2} after electroweak symmetry breaking; the former dominates for  $m_{\ntilde}/\hv\textgreater 10$. We will consider both $m_{\ntilde}\,\textgreater\,m_\psi$ and $m_{\ntilde}\,\textless\,m_\psi$ in this paper. In Fig.\,\ref{fig:spectrum}, we show the mass spectrum and the decays relevant for our study in the $m_{\ntilde}\,\textgreater\,m_\psi$ scenario. Finally, $N_2,\,N_3,\,\tilde{N}_2,$ and $\tilde{N_3}$ decay via the mixings with their active neutrino or sneutrino counterparts. 

\section{Formalism}
\label{sec:formalism}

The goal of this paper is to highlight new qualitative features arising in the supersymmetric framework. We focus on scenarios where $\phi$ is in equilibrium at high temperatures $T\,\textgreater\, m_\phi$, and its decays during this period result in the freeze-in production of $N_1$ and $\ntilde$. No significant production of $N_1$ or $\ntilde$ occurs after $\phi$ freezes out, as it decays rapidly into lighter SM or SUSY particles. For cases where $m_{\ntilde}\,\textless\,m_\psi$, we take $m_\psi\gg m_\phi$ so that $\phi$ decays remain the dominant source of $\ntilde$ and $N_1$ production. We ignore the cases where $\phi$ itself freezes in, which can also produce sterile neutrino dark matter \cite{Merle:2013wta,Adulpravitchai:2014xna,Kang:2014cia,Roland:2014vba,Merle:2015oja}, or where $\psi$ decay is the dominant production mechanism, since they do not demonstrate any qualitatively new features.

The conditions that $\phi$ maintain equilibrium with the thermal bath while $N_1$ and $\ntilde$ both freeze in from $\phi$ decays enforce the following relations between couplings and masses \cite{Hall:2009bx} (we simplify $x\equiv x_1, A_\phi\equiv {A_x}_1$):
\be
\lambda^2\,\textgreater \frac{m_\phi}{M_{\text{Pl}}},~~~x^2\, \textless\, \frac{m_\phi}{M_{\text{Pl}}},~~~x^2\,\frac{A_\phi^2}{m_\phi^2}\, \textless\, \frac{m_\phi}{M_{\text{Pl}}}.
\label{eq:freezeincondition}
\ee
Crucially, note that this feeble coupling $x\ll 1$ results in a long lifetime for $\ntilde$.

Since each $\ntilde$ decay produces an $N_1$ particle, both direct $\phi$ decays and $\ntilde$ decays contribute to the final dark matter population; these contributions are approximately \cite{Hall:2009bx}:
\bea
\Omega_{N_1} h^2(\phi) &\sim& \frac{10^{24}\, x^2}{2\pi\,S_{N_{2,3}}}\frac{m_{N_1}}{m_\phi}\label{rd1}\\
\Omega_{N_1} h^2(\tilde{N}_1) &\sim& \frac{10^{24}\, x^2}{2\pi\,S_{N_{2,3}}}\frac{m_{N_1}}{m_\phi}\left(\frac{A_\phi}{m_\phi}\right)^2\label{rd2}
\eea
Here, $S_{N_{2,3}}$ ($\sim 1-30$ for GeV scale $N_2,N_3$ \cite{Scherrer:1984fd,Bezrukov:2009th,Asaka:2006ek}) accounts for entropy dilution from the late freeze-out and out-of-equilibrium decay of the other two sterile neutrinos $N_2,N_3$. 

Since $\ntilde$ decays produce Higgsinos, we must ensure that $\ntilde$ decays before Higgsino freeze-out in order for $N_1$ to form the dominant DM component. Using the radiation-dominated time-temperature relation $H(T) = T^2/M_0$ with $M_0 = \left(\frac{45 M_{Pl}^2}{4\pi^3 g_*}\right)^{1/2}$, where $g_*$ is the number of degrees of freedom in the bath, the temperature of the SM bath when $\ntilde$ decays is approximately $T_{\text{decay}}\approx (\Gamma_{\ntilde} M_0)^{1/2}$, where $\Gamma_{\ntilde}$ is the decay width of $\ntilde$. In our calculations, we ensure that $T_{\text{decay}}$ is higher than the Higgsino freeze-out temperature $\sim m_{\tilde{H}}/20$. 

Sterile neutrino dark matter can be cold, warm, or hot, as characterized by its free-streaming length $\Lambda_{FS}$, defined as the distance travelled by a dark matter particle from its production at time $t_p$ to the present time $t_0$
\be
\Lambda_{FS} = \int_{t_p}^{t_0} \frac{\langle v(t) \rangle}{a(t)} \diff t\,.
\ee
Here $v(t)$ and $a(t)$ are the DM velocity and the scale factor respectively at a given time t. As a rough guide, we take $\Lambda_{FS}\lesssim0.01$ Mpc, $0.01\lesssim \Lambda_{FS} \lesssim 0.1$ Mpc, and $0.1 ~\text{Mpc}\lesssim\Lambda_{FS}$ as corresponding to cold, warm, and hot dark matter respectively \cite{Merle:2015oja}.

If $m_{\ntilde}\gg m_{N_1}$ and $\ntilde$ decays extremely late, the population of $N_1$ produced from such decays can be relativistic and act as dark radiation. It is well known that a species that forms all of dark matter cannot account for any measurable dark radiation in the Universe \cite{Hasenkamp:2012ii,Merle:2015oja,Reece:2015lch}. However, this constraint can be circumvented in our framework since the hot $N_1$ population produced from $\ntilde$ decays does not mix with the cold $N_1$ population from $\phi$ decays. The latter population can thus be the dominant dark matter component, while a subdominant, hot population from $\ntilde$ decays forms dark radiation; we conservatively take this fraction to be $\lesssim 1\%$ (as in \cite{Hooper:2011aj}), which should leave structure formation unaffected. We note that heavy, long-lived $\ntilde$ can grow to dominate the energy density of the Universe, introducing an intermediate phase of matter domination, subsequently releasing entropy that reheats the thermal bath and dilutes the dark matter abundance. This indeed occurs in parts of our parameter space, and we correct for these effects appropriately.

Such energetic $N_1$ contribute to the effective number of relativistic degrees of freedom $\dn$ during Big Bang Nucleosynthesis (BBN) (which we take to be at $T_{BBN}=4$ MeV). This contribution can be estimated as
\begin{equation}
\dn=\left.\frac{\rho_{N_1}}{\rho_{\nu}}\right|_{T=T_{BBN}},
\end{equation}
which compares the sterile neutrino energy density with the energy density of a neutrino species in equilibrium at the same temperature. Current bounds on $\dn$ at BBN are at the level of $\sim 0.3$ at 1$\sigma$ \cite{Cyburt:2015mya}. With the simplifying assumption that all of the $\ntilde$ population decays at $T_{decay}$ and $N_1$ is produced with typical energy $\mntilde/2$ ($\mntilde/3$) in a two (three) body decay process, which gets redshifted by a factor ${S_{N_{2,3}}^{1/3} ({g_*}_{SM}/{g_*}_{BBN})^{1/3}}$ due to subsequent entropy dilution, $\dn$ can be approximated as (for the three body decay case)
\bea
&&\dn\approx\frac{10^{-8}}{S_{N_{2,3}}^{1/3} ({g_*}_{SM}/{g_*}_{BBN})^{1/3}}\,\Omega h^2 \frac{m_{\tilde{N}_1}}{T_{decay}}\frac{\text{GeV}}{m_{N_1}}\nonumber\\
&\approx& 0.2\left(\frac{\Omega h^2}{0.0012}\right) \left(10^{-8}\frac{m_{\tilde{N}_1}}{T_{decay}}\right)\left(\frac{\text{MeV}}{m_{N_1}}\right)\left(\frac{10}{S_{N_{2,3}}}\right)^{1/3}
\label{neffeqn}
\eea
Here, $\Omega h^2$ represents the present relic abundance that originated from $\ntilde$ decay, as this is the only component that is relativistic at BBN. 

While there are stronger constraints on $\dn$ from the later era of Cosmic Microwave Background (CMB) decoupling, the $N_1$ particles generally redshift and become nonrelativistic by this time \cite{Merle:2015oja}, resulting in weaker constraints, hence we only focus on $\dn$ during BBN. However, we do note that light (sub-eV) mass sterile neutrinos produced in this manner could contribute to $\dn$ at CMB decoupling, and might be relevant for alleviating the recent tension between the local and CMB-inferred measurements of the Hubble rate \cite{Bernal:2016gxb}. 

\section{Results}

In this section, we investigate modifications to dark matter properties in the supersymmetric framework. 

\textit{Abundance and Composition:} The $N_1$ population acts as multi-component dark matter as the fractions produced from $\phi$ and $\ntilde$ decays do not interact with each other. The two abundances differ by a factor of $(A_\phi/m_\phi)^2$ (see Eqs.\,\ref{rd1},\,\ref{rd2}). Since we expect $A_{\phi}\sim m_\phi\sim m_{\text{SUSY}}$, the two abundances are generally of comparable magnitude. For given values of $m_\phi$ and $m_{N_1}$, the desired relic abundance can be obtained by selecting appropriate values of $x$ and $A_\phi$ as long as Eq.\,\ref{eq:freezeincondition} remains satisfied. Due to the presence of an additional dark matter production mechanism in $\ntilde$ decays, the supersymmetric framework opens up more parameter space where sterile neutrino dark matter can be realized.

\begin{figure}[t!]
\includegraphics[width=3.2in]{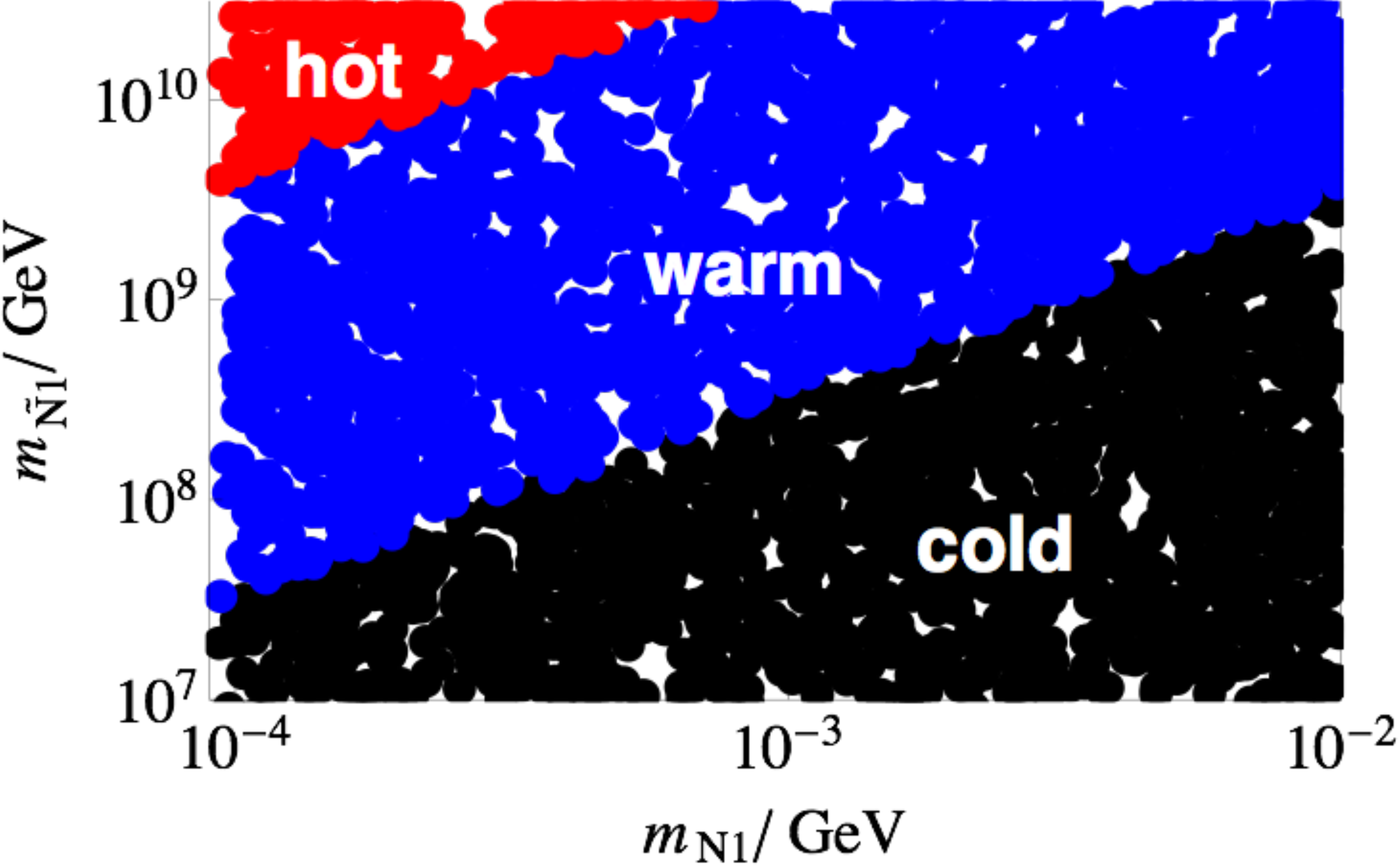}
\caption{\label{fig:fslength1} Parameter space with cold, warm, and hot dark matter (black, blue, and red regions respectively). For all points in the plot, $\Omega h^2=0.12$, $m_\phi=10^{11}$ GeV, $A_\phi/m_\phi= 10$, $S_{N_{2,3}}=10$.}
\end{figure}

\begin{figure}[t!]
\includegraphics[width=3.2in]{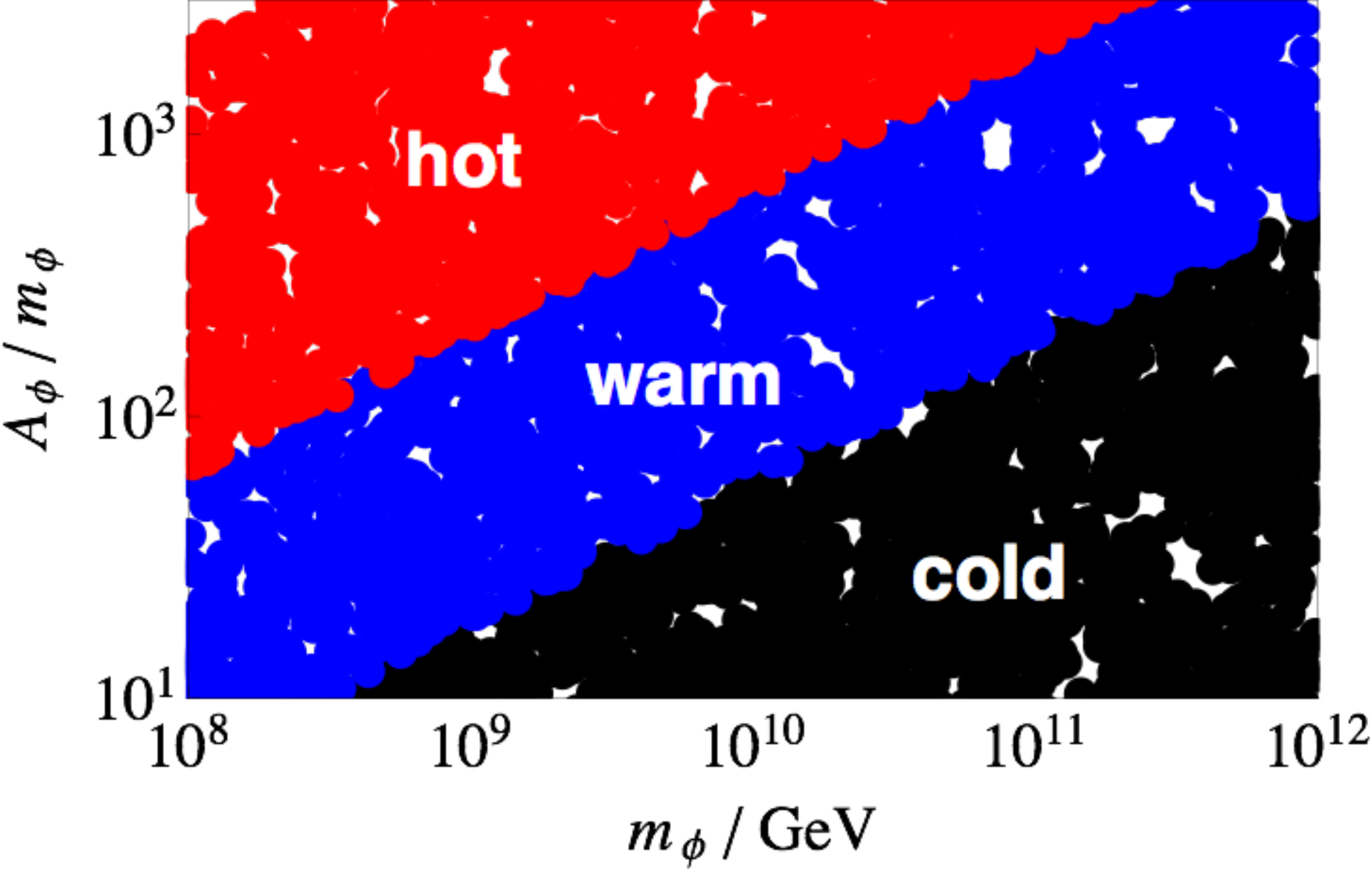}
\caption{\label{fig:fslength2} Cold, warm, hot dark matter (black, blue, and red regions respectively) for $m_{N_1}=1$ MeV and $m_{\ntilde}=10^6$ GeV. We set $S_{N_{2,3}}=10$.}
\end{figure}

\textit{Free-streaming length:} $\ntilde$ decays can produce dark matter that is cold, warm, or hot. This is illustrated in Fig.\,\ref{fig:fslength1}, where we delineate combinations of sterile neutrino and sterile sneutrino masses that give rise to cold, warm, or hot dark matter (regions where the full dark matter relic density can be achieved extend beyond the boundaries of this plot). In this plot, $m_{\ntilde}\,\textgreater\, m_\psi$, so that $\ntilde$ decays as $\ntilde\rightarrow\psi N_1$; $m_\phi=10^{11}$ GeV, so that $\phi\rightarrow\ntilde\ntilde$ is allowed at all points; $A_\phi\,=\,10\,m_\phi$, so that $\ntilde$ decays account for essentially all of dark matter; and $x$ is chosen to produce the desired relic density $\Omega h^2=0.12$. As expected, heavier $\ntilde$ or lighter $N_1$ cause dark matter particles to become more energetic, resulting in larger free streaming lengths. Note, however, that the demarcation of cold, warm, and hot regions depends not only on $\mntilde$ and $\mn$ but also on other parameters (in particular, the ones that determine the $\ntilde$ lifetime); this point is illustrated in Fig.\,\ref{fig:fslength2}, where we show that all three possibilities can be realized for the same choice of $\mntilde$ and $\mn$ (fixed to $10^6$ GeV and $1$ MeV respectively) by varying $m_\phi$ and $A_\phi$. 

\textit{Dark radiation:}
Next, we consider scenarios where extremely energetic $N_1$ from late $\ntilde$ decays contribute significantly to $\dn$ during BBN. Here we choose $m_{\ntilde}\,\textless\, m_\psi$ so that $\ntilde$ decays through the three-body channel $\ntilde\rightarrow N_1 \tilde{H} h$ with a long lifetime. As discussed in the previous section, this $N_1$ population can only comprise a subdominant component of dark matter, and we fix its abundance to $1\%$ of the total DM abundance by choosing $A_\phi=0.1\, m_\phi$. 

\begin{figure}[t!]
\includegraphics[width=3.65in]{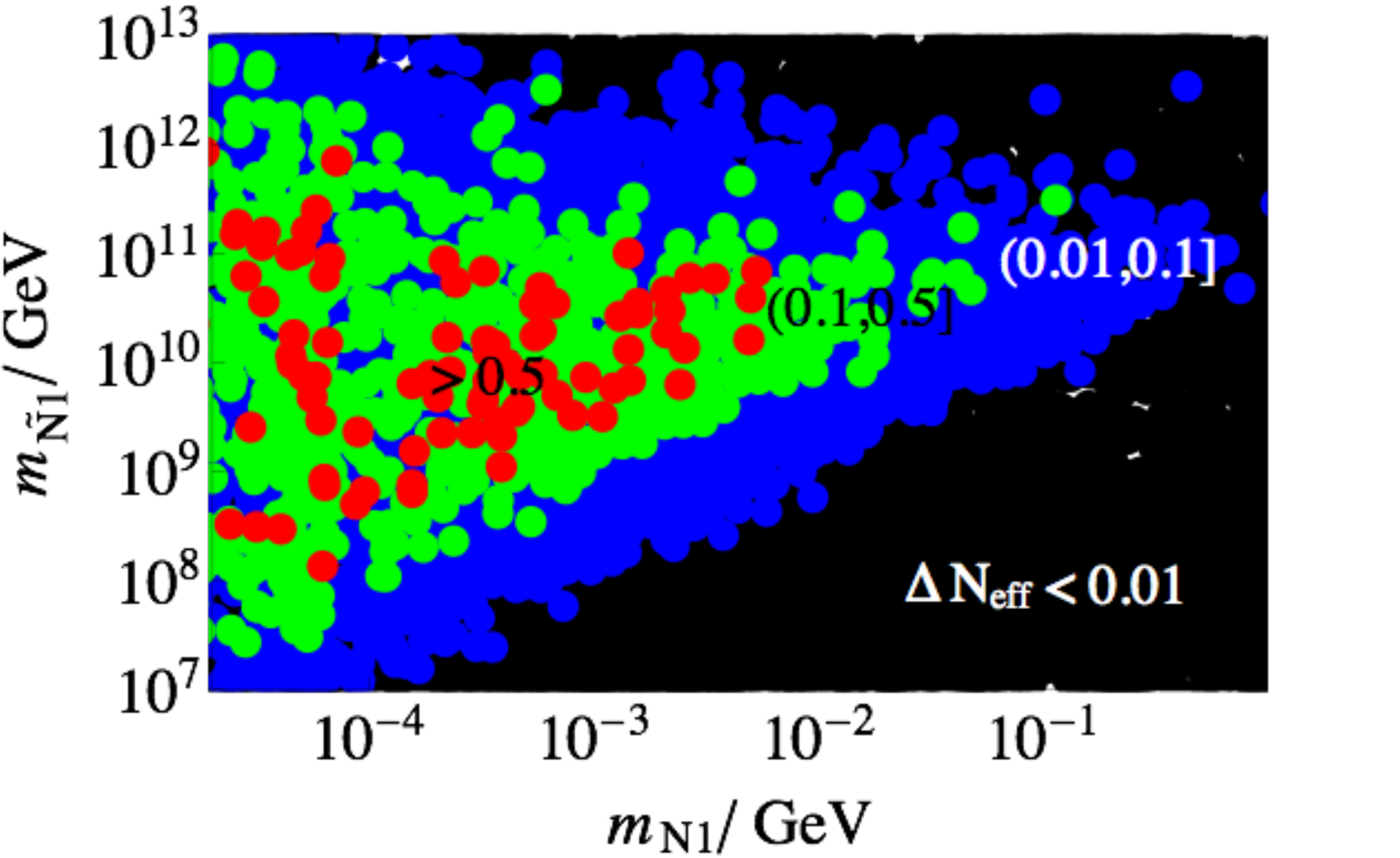}
\caption{\label{fig:neff} $\dn$ (BBN) for different $N_1$ and $\ntilde$ masses. Red, green, blue, and black points denote $\dn$ in the ranges $\textgreater\, 0.5,\, 0.1-0.5,\, 0.01-0.1$, and $\textless\, 0.01$ respectively. For all points, the $\dn$ contribution comes from $\ntilde$ decays, which account for $1\%$ of the dark matter abundance, while $\phi$ decays produce the rest of dark matter.}
\end{figure}

In Fig.\,\ref{fig:neff} we plot $\dn$ at BBN as a function of $N_1$ and $\ntilde$ masses from a scan over parameter space, where we scanned over $S_{N_{2,3}}=1-30$. Red, green, blue, and black points represent $\dn$ in the ranges $\textgreater \,0.5,\, 0.1-0.5,\, 0.01-0.1$, and $\textless\, 0.01$ respectively; we see that large contributions to $\dn$ comparable to current bounds are possible while satisfying all the enforced constraints. The largest values correspond to $\mn\sim$ MeV and $\mntilde\sim 10^{9}-10^{12}$ GeV: for lighter $\ntilde$ or heavier $N_1$, the DM particles are not sufficiently relativistic at BBN, whereas heavier $\ntilde$ (which forces $\phi$ to be heavier) or lighter $N_1$ both require larger $x$ to maintain the correct dark matter abundance (see Eq.\ref{rd1}), which reduces the $\ntilde$ lifetime.  

\section{Discussion}

In this paper, we have demonstrated that a supersymmetric extension of the widely studied sterile neutrino dark matter framework with the basic features of dark matter freeze-in, namely an underlying symmetry that stabilizes the dark matter candidate and a feeble coupling that facilitates dark matter production, can 
introduce several qualitatively new cosmological features and dark matter properties that are not possible in the non-supersymmetric scenario.  

The presence of the superpartner, the sterile sneutrino $\ntilde$, offers an additional production mechanism for dark matter. In addition to extending the allowed parameter space for sterile neutrino dark matter, this makes possible the scenario of multiple-component dark matter with a single constituent $N_1$, as the fractions produced via different processes ($\phi$ and $\ntilde$ decays) do not mix, and effectively act as different components. Note that this possibility is unique to freeze-in production, as the two fractions would thermalize in the standard dark matter (freeze-out) histories if such production occurred before freeze-out. $\ntilde$ decays can be the dominant source of dark matter production, and dark matter produced via its decay can be cold, warm, or hot. The scenario of mixed dark matter (some combination of cold and warm components) might hold interesting implications for structure formation, offering resolution to issues such as the core vs.\ cusp problem and the ``too big to fail" problem \cite{Lovell:2011rd, BoylanKolchin:2011dk}. In scenarios where $\ntilde$ is long lived and produces $\sim 1\%$ of the dark matter population, $\mathcal{O}(0.1)$ contributions to the effective number of relativistic degrees of freedom $\dn$ during BBN are possible, which can be probed by near future measurements. Similar results were also discussed in \cite{Roland:2016gli}, but within a constrained framework that only offered limited possibilities. Such mixed dark matter scenarios with a single particle constituent and the associated dark matter phenomenology deserve more attention and careful study. 

There are several interesting supersymmetric model building aspects that we have not fully addressed. The $\mathcal{Z}_2$ symmetry that makes the vanishing mixing of $N_1$ technically natural can be embedded into the details of the underlying theory. This $\mathcal{Z}_2$ symmetry also need not be exact, in which case $N_1$ can decay. This prospect is especially appealing given the recent claims of an X-ray line from galaxy clusters at 3.5 keV \cite{Bulbul:2014sua,Boyarsky:2014jta} compatible with decays of a 7 keV sterile neutrino; this direction would warrant further study should the signal persist. Likewise, the most interesting regions of parameter space from a phenomenological point of view require a large hierarchy between $\ntilde$ and $N_1$ masses; these could emerge naturally from symmetry considerations in the supersymmetric neutrino sector, see $\textit{e.g.}$ \cite{Roland:2014vba,Roland:2015yoa}.

Given the tremendous appeal of supersymmetry as part of the underlying theory of nature, the cosmological aspects discussed in this paper are relevant for any study on sterile neutrino dark matter. Moreover, in the absence of clear observational signals of weak scale supersymmetry or WIMP dark matter, such lines of inquiry might provide hints on the nature and scale of supersymmetry and open up promising avenues of research in the future.

\medskip
\textit{Acknowledgements: } We acknowledge helpful discussions with Samuel Roland. The authors are supported in part by the DoE under grants DE-SC0007859 and DE-SC0011719. BS also acknowledges support from the University of Cincinnati. JW wishes to acknowledge support from the Humboldt Foundation. This work was performed in part at the Aspen Center for Physics, which is supported by National Science Foundation grant PHY-1066293.
\bibliography{neutrino_bibliography}

\end{document}